\documentclass[epj,referee]{svjour}
\usepackage{latexsym}
\usepackage{graphics}
\usepackage{epsfig}

\begin{document}
\newcommand{\lco} {La$_2$CuO$_4$}
\newcommand{\lsco} {La$_{2-x}$Sr$_x$CuO$_4$}
\newcommand{\lscotu} {La$_{1.98}$Sr$_{0.02}$CuO$_4$}
\newcommand{\lscofri} {La$_{1.97}$Sr$_{0.03}$CuO$_4$}
\newcommand{\la} {$^{139}$La}
\newcommand{\cu} {$^{63}$Cu}
\newcommand{\cuo} {CuO$_2$}
\newcommand{\ybco} {YBa$_{2}$Cu$_3$O$_{6.1}$}
\newcommand{\ybcoca} {Y$_{1-x}$Ca$_x$Ba$_2$Cu$_3$O$_{6.1}$}
\newcommand{\yt} {$^{89}$Y}
\newcommand{\etal} {{\it et al.}}
\newcommand{\ie} {{\it i.e.}}
\title{
Anelastic relaxation and \la NQR in \lsco around the critical 
Sr content x=0.02
}
\author{A.Campana\inst{1}, M.Corti\inst{1}, A. Rigamonti\inst{1} \and
R.Cantelli\inst{2} \and F.Cordero\inst{3}} 
\institute{Dipartimento di Fisica ``A. Volta'' e Unit\'a INFM di Pavia,
Via Bassi 6, 27100 Pavia, Italy
\and Department of Physics, University of Roma ``La Sapienza'' and Unit\'a 
INFM P.zale A.Moro 2, I-00185, Roma (Italy)
\and CNR, Area di Ricerca di Roma - Tor Vergata, Istituto di Acustica ``O.M. Corbino'',
Via del Fosso del Cavaliere, I-00133, Roma and INFM (Italy)}
\date{Received: date / Revised version: date}
%
\abstract{
Anelastic relaxation and \la \hspace{0.3cm} NQR relaxation measurements in \lsco,
for Sr content x around 2 and 3 percent, are presented and discussed in terms
 of spin and lattice excitations and ordering processes. It is discussed how 
the phase diagram of \lsco at the boundary between the antiferromagnetic
 (AF) and the spin-glass phase (x = 0.02) could be more complicate than previously 
thought, with a transition to a quasi-long range ordered state at     
  T = 150 K, as indicated by recent neutron scattering data. On the other hand, 
the \la  NQR spectra are compatible with a transition to a conventional AF phase 
around T = 50 K, in agreement with the phase diagram commonly accepted in the 
literature. In this case the relaxation data, with a peak of magnetic origin in the NQR 
relaxation rate around 150 K at 12 MHz and the anelastic counterparts around
 80 K in the kHz range, yield the first evidence in \lscotu of 
freezing involving simultaneously lattice and spin excitations. 
This excitation could correspond to the motion of charged stripes.
}
%
\titlerunning{Anelastic relaxation and \la NQR in \lsco...}
\maketitle
	\section{Introduction}
\label{intro}
From a variety of recent experiments and theoretical descriptions (mostly motivated by 
the search of the microscopic mechanism underlying high-temperature superconductivity),
 it has been realized that the electron system in doped two-dimensional (2D) quantum 
Heisenberg antiferromagnets (AF) exhibits complicated ordering phenomena. On cooling 
from high temperatures, first a kind of phase separation is expected to occur, causing 
the formation of charged stripes separating mesoscopic AF domains \cite{stripes,zaanen,emery}. 
In cuprates
in general the stripes should exist only dynamically, with slowing down of their 
fluctuations on cooling. At lower temperatures the spin degrees of freedom associated 
to the AF patches \cite{gooding} between the stripes are known to freeze, generating a cluster 
spin-glass state \cite{weid,chou,nieder}. 
Charge and spin freezing both involve a complex spin dynamics. 
While the stripe dynamic is slow, the motion of the holes along the stripe is much
faster than the fluctuation of the stripe itself. 


There is also evidence of unusual coupling of the lattice to charge and spin
excitations \cite{egami}. 
For instance, the \la NMR line broadening, for T $\leq$ 40 K, in
\lsco \\
(LSCO) for x = 0.12 (a signature of modulated magnetic order) 
is accompanied by softening of sound velocity \cite{suzuki}. Neutron diffraction, for 
0$\leq$ x $\leq$ 0.3, indicates local tilts of octahedra, interpreted as evidence 
of charged stripes \cite{bozin}, the local tilt decreasing with increasing x. The 
local tilts give also rise to tunneling systems, observed by acoustic 
experiments for x $\leq$ 0.03, and relaxation rate strongly depends on doping \cite{cordero}.


Charge localization along stripes and spin freezing have been studied, in 
\lco-based compounds, mostly by means of NMR-NQR and $\mu$SR spectroscopies, 
which probe the low frequency excitations through the relaxation times and
the modifications in the spectra \cite{borsa}. In particular, it has been argued
\cite{hunt} that,
in the underdoped regime of LSCO, when diffraction experiments indicate
complete ordering, the stripes are still fluctuating at low frequencies.
LSCO at Sr content around 0.02 is interesting, being at the boundary 
between the 3D-AF and the spin-glass phase (see Ref. \cite{john} for a review). 
A recent neutron scattering study has shown that quasi-3D magnetic
ordering occurs below about 40 K in the spin-glass state, with a spin 
structure related to the diagonal stripe structure \cite{matsuda}. This new intermediate
magnetic state is believed to result from partial freezing of 2D spin
fluctuations existing at high temperatures, the spin-glass state being 
described as a random freezing of quasi-3D spin clusters with anisotropic 
spin correlations.


Motivated by this scenario of interrelated lattice and spin fluctuation
effects, we have undertaken a comparative study of LSCO at x = 0.02 and
x = 0.03 based on anelastic relaxation, \la NQR relaxation and NQR spectra.


We first qualitatively recall how slowing down of spin fluctuations and 
ordering are expected to affect nuclear and mechanical relaxation. Single 
holes or charged stripes motions cause a time dependence in the hyperfine
field {\bf h}(t) =$\sum_i$ {\bf A}$_i${\bf S}$_i$(t) 
at the nucleus ({\bf S}$_i$ spin operator at the {\it i}-th ion,
{\bf A}$_i$ hyperfine coupling tensor). When a characteristic frequency $\omega_s$ of the
fluctuating stripes becomes of the order of the measuring quadrupole frequency
$\omega_m$, a maximum in the spin lattice relaxation rate T$_1^{-1}$ = $W$ driven by the local
time dependence of {\bf h}(t) is expected. Below this temperature the stripes move 
very slowly, or are "pinned", and an "anomalous" magnetic moment is induced,
associated to the 2D patches of AF correlated ions in between stripes \cite{gooding}. The
randomly distributed magnetic moments $\mu_i$ experience cooperative slowing down
and a second relaxation mechanism sets in, related to the field at the nuclear 
sites due to $\mu_i$'s. In the correspondent relaxation rate a correlation function
of the form $\langle \sum_{i,j} \mu_i(0) \mu_j(t)\rangle = 
\sum_i \langle \mu_i(0) \mu_i(t)\rangle$ is involved.


One can empirically
write this correlation function as $\mu^2$ exp[-t/$\tau_f(x,T)$], with an average correlation
time $\tau_f$ which increases on decreasing temperature. At the temperature T$_g$ where 
$\tau_f$ becomes of the order of $\omega_m^{-1}$ another peak appears in T$_1^{-1}$, 
with non-exponential 
recovery law, a signature of disordered systems. Below T$_g$ one speaks of spin freezing.
In a long-range ordered AF matrix T$_g$ should increase about linearly with x, due to
the increased strength of the interaction among the $\mu_i$'s. On the contrary, for an 
amount of doping x which destroys the long range AF order, with the onset of a 
cluster spin-glass phase, T$_g$ decreases with increasing x \cite{borsa,cho,carretta}.


In particular, in LSCO, for x $\leq$ 0.02 the peaks in \la NQR {\it W} have been shown
\cite{chou}
to follow the law T$_g$ = T$_f$ = bx, indicating spin freezing in an AF matrix. 
For x $\geq$ 0.02 (cluster spin-glass phase) the peaks at T$_g$(x) have been attributed 
to the spin freezing of the magnetic moments in short-range AF correlated islands
\cite{cho}.


The stripe localization has been detected indirectly from the wipe-out 
effect on the \cu  NQR signal in Nd and Eu doped LSCO at x = 0.12, at
T$_{charge}$ = 65 K (and in the underdoped regime of LSCO), with a wipe 
out fraction having a temperature behavior similar to the charge and
spin order determined by neutron scattering \cite{hunt,singer}. It should be remarked
that in principle a wipe out effect should be accompanied, at higher
temperature, by a marked enhancement in the spin lattice relaxation 
rate, when $\omega_s$ = $\omega_m$. Finally, $\mu$SR and \la  NQR measurements 
\cite{nieder,julien} pointed
out a magnetic transition to a spin-glass like phase well extending 
into the superconducting regime.


As regards the effects expected in the anelastic relaxation, one notes 
that the elastic energy loss coefficient Q$^{-1}$, measured by exciting 
flexural vibrations, is directly proportional to the imaginary part of 
the mechanical susceptibility $\chi"(\omega)$. Thus Q$^{-1}$ is related to the spectral 
density J$_{latt}(\omega)$  of the motions causing dissipation, according to the law 
Q$^{-1}\propto \chi"\propto \omega J_{latt}$.


Since the motions of the stripes involve sizeable lattice effects, when a
characteristic frequency decreases down into the kHz range they can be 
detected as maxima in $\chi"/\omega$. Thus, from a combination of anelastic
relaxation and of magnetic NQR relaxation, one can in principle probe 
the lattice and the spin fluctuations associated to the stripe motions.
The investigation reported here was aimed at this purpose.
\section{Experimentals  results and discussion}
\label{sec:1}
Two LSCO ceramic samples grown by standard solid state reaction \cite{ferr} 
have been investigated. 
According to x-ray diffraction the final amounts were x = 0.022 and x = 0.032 respectively.
A more precise estimate of x was derived by detecting the orthorhombic-tetragonal
transition through anelastic relaxation. The relationship of the transition
temperature T$_0$ to the amount of Sr was taken as T$_0$ (x) = 535 [1 - (1/0.235) x] 
(Refs. \cite{john} and \cite{cho2}). 
From the step in the Young's modulus 
(Fig.1) at the transition 
normalized in temperature and amplitude, one as T$_0$ = 495 K and T$_0$ = 468 K, 
corresponding to the Sr amounts x = 0.019$\pm$0.0015 and x=0.030$\pm$0.001 (hereafter
called samples 2 and 3 percent). 
As it is noted the structural transition
appears even sharper than in pure \lco (where some broadening may be
attributed to non-perfect oxygen stoichiometry).
\begin{figure}
\resizebox{0.7\textwidth}{!}{%
\includegraphics{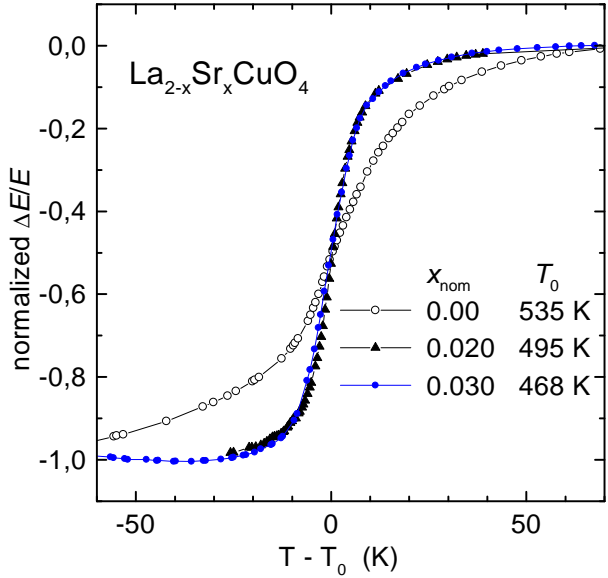}
}
\vspace{-1.5cm}       
\caption{Variation of the Young modulus at the tetragonal orthorhombic transitions,  
normalized in amplitude and reported as a function of (T-T$_o$).}
\label{fig:1}       
\vspace{-0.6cm} 
\end{figure}
SQUID magnetization 
measurements yield a magnetic susceptibility as a function of temperature
qualitatively typical of a spin-glass phase. Small differences between the
field cooled and zero- field cooled data below about 140 K have been 
attributed to magnetic impurities present in the powders used for the 
preparation. These impurities do not affect the \la NQR and anelastic 
relaxation measurements.


In Fig.2 the quantity J$_{latt}$ = T Q$^{-1}/\omega$  is reported in the temperature 
range of interest, in correspondence to the measuring frequencies
$\omega/2\pi$ = 1.29, 6.9 and 17.2 kHz, in LSCO at x = 2 percent. The
thermal depinning of the stripes should have cooperative character. 
For the moment we neglect their frequency distribution. According to
the data in Fig. 2, the spectral density of the motion responsible 
of the dissipation has a diffusive character: 
 \begin{equation}
  \label{gei}
  J_{latt}(\omega_m) = \frac{2 \omega_s}{\omega_s^2 + \omega_m^2}
 \end{equation}
 \begin{figure}
  \resizebox{0.7\textwidth}{!}{%
  \includegraphics{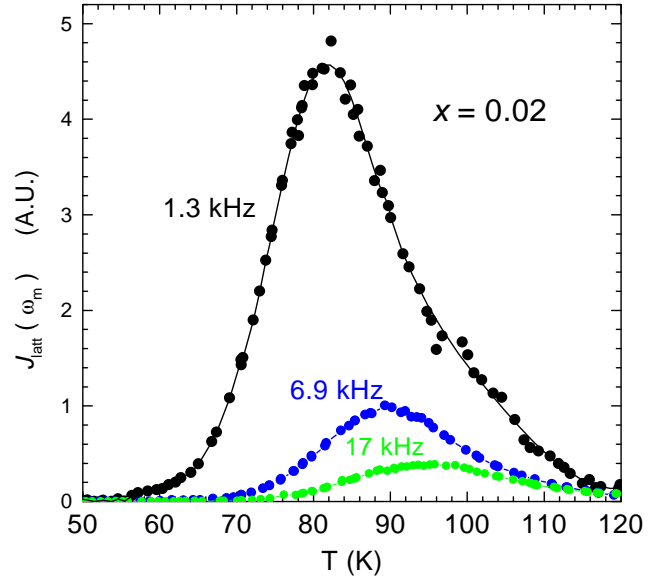}
  }
  \vspace{-1.5cm}       
  \caption{Spectral density of the lattice motion responsable of the elastic energy 
  loss coefficient in LSCO at Sr content x=0.02 as a function
  of temperature for three  measuring frequencies.}
  \label{fig:2}       
 \end{figure}
From the temperature where the maxima are observed one can deduce
the values of the characteristic frequency $\omega_s$. A good fit of the 
data (Fig.3) is obtained on the basis of a temperature behavior of
$\omega_s$ of the form
 \begin{equation}
  \label{om}
  \omega_s  = \omega_0 exp(-E/T)	]
 \end{equation}
with $\omega_0$ = 4.5$\times$10$^{12}$ s$^{-1}$ and E = 1650 K, consistent with the idea of
thermal activation of the stripe fluctuations.
 \begin{figure}
  \resizebox{0.7\textwidth}{!}{%
  \includegraphics{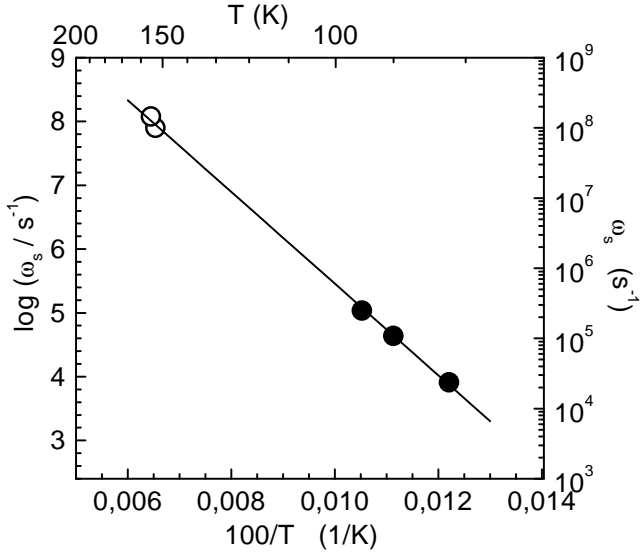}
  }
  \vspace{-1.5cm}       
  \caption{Characteristic frequencies  of the motions as estimated from the
  maxima in Fig. 2. The solid line is the fitting behavior according to Eq.2 in
  the text and  correspondent to  $\omega_0$ = 4.5 1012 s$^{-1}$    
  and  E=1650 K. The empty circle 
  indicate the frequencies deduced 
  from the maxima in the \la NQR relaxation rate around T=160 K  
  ( see Fig. 5 and text).}
  \label{fig:3}       
 \end{figure}
As regards \la NQR relaxation, the recovery laws for 
the +5/2 - +7/2 line at 3$\nu_Q$ and the +3/2 - +5/2 at 2$\nu_Q$, 
are multiexponential. However in the first decade they differ 
only little from a single exponential. The correspondent
effective decay rate $\tau_e^{-1}$ can be related to the magnetic
relaxation rate $W_M$ or to the quadrupolar relaxation rate $W_Q$ 
due to the time dependence of the electric field gradients at 
the La site, in the following way \cite{rega}:
 \begin{eqnarray}
  \label{rec}
  3\nu_Q  \hspace{0.5cm}
  \tau_e^{-1} &=& (67/21) W_Q = 23 W_M \nonumber	\\				
  2\nu_Q  \hspace{0.5cm}
  \tau_e^{-1} &=& (64.5/21) W_Q = 41.3 W_M
 \end{eqnarray}
For T $\geq$ 250 K the NQR relaxation rate (Fig.4) 
are frequency independent and follow the law $\tau_e^{-1}\propto T^2$, 
features characteristic of the relaxation process driven
by underdamped phonons \cite{adv}. Also an order of magnitude
estimate corroborates this conclusion, since for this
process one expects \cite{birge} $W_Q$ = 5$\times$ 10$^{-4}$ T$^2$ s$^{-1}$. 
Around 280 K 
some evidence of a quadrupole contribution due to overdamped
phonon modes (tilting of the oxygen octahedra in a double
well potential \cite{prb}) is present \cite{rr}. Below 250 K the relaxation 
rates depart from the behavior described above. For temperatures
lower than about 180 K the comparison of the data for the 3$\nu_Q$
and 2$\nu_Q$ lines indicates the insurgence of a magnetic relaxation
mechanism. The temperature behavior of $\tau_e^{-1}$ for T $\leq$ 230 K is
analyzed in detail in Fig. 5, after subtraction of the background 
of quadrupole character. Assuming that the modulation of the
hyperfine magnetic field {\bf h}(t) is due to the same motion
causing the mechanical relaxation, then for
 \begin{equation}
  \label{2vu}
  2W_M = \frac{1}{2} \gamma^2 \int{\langle h_+(0)h_-(t)\rangle e^{-i\omega t} dt}
 \end{equation}
one writes 
 \begin{equation}
  \label{tau}
  (\tau_e^{-1})_{2nQ/3nQ} = aW_M = a\frac{1}{2}\gamma^2  h^| [2\omega_s /(\omega_s^2 + 
  \omega_m^2)]	
 \end{equation}
with a = 23 for the 3$\nu_Q$ line and a = 41.3 for the 2$\nu_Q$ line, $\omega_m$ = 3$\omega_Q$ 
and $\omega_m$ = 2$\omega_Q$ respectively. 
In Fig. 5 the experimental data for $aW_M$ are
compared with the theoretical behaviors for the relaxation rates
according to Eq.s \ref{rec} and \ref{tau}, having used for $\omega_s$ the expression
derived from the anelastic relaxation (Eq. \ref{om}). The maxima in $W_M$ 
are well reproduced. The departures of the experimental data from
the theoretical expressions in the temperature range corresponding 
to slow motions, {\it i.e.}  
$\omega_s \leq \omega_m$, are likely to be due to the simplifying
assumption of a monodispersive process. A distribution in $\omega_s$ implies
a flattening in the relaxation rate around the maximum and a departure
from the behavior for monodispersive process more marked in the low
temperature range, as it is observed in the Figure. The relevant fact
is that the temperature dependence of $\omega_s$ deduced from anelastic 
relaxation in the kHz range seems to justify quantitatively the
magnetic NQR relaxation rate in the MHz range.
 \begin{figure}
  \resizebox{0.62\textwidth}{!}{%
  \includegraphics{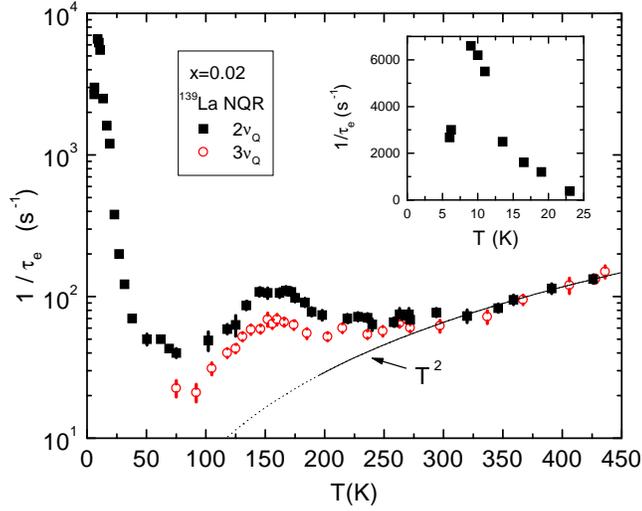}
  }
  \vspace{-1.5cm}       
  \caption{Effective decay rate in the recovery plots for 139 La NQR relaxation 
  in LASCO at x=0.02 , for the 3$\nu_Q$ line  ($\circ$) and for the 2$\nu_Q$ line ($\Box$).
  The inset 
  is the blow up of the data for T$\leq$25 K. The solid line is the sketch of the 
  temperature dependence for relaxation driven by underdamped  phonons.}
  \label{fig:4}       
  \vspace{-0.6cm} 
 \end{figure}
On the other hand, the maxima in $W_M$ around 150 K could reflect the 
slowing down of the spin dynamics on approaching the transition to 
an 
ordered state. Information in this regard can be obtained from the 
NQR spectra (Fig.6), since the insurgence of a static field $\langle h\rangle$ at
the La site is signaled by a splitting $\Delta$ of the resonance line,
proportional to the sublattice magnetization $\langle S \rangle$. A clear splitting 
of the line is noticeable only below about 50 K, close to the temperature
indicated by neutron scattering \cite{matsuda}.  If the width $\delta$ 
of a single component 
is kept temperature independent the fitting of the spectra with two 
gaussian lines yields an ordering temperature T$_N$, where $\Delta$ goes to zero, 
of about 140 K. The temperature dependence of the order parameter 
$\langle h\rangle\propto \langle S\rangle$ 
would turn out quite different from the one of a canonical phase transition 
to the AF phase, experimentally observed \cite{macl} in pure \lco.
 \begin{figure}
  \resizebox{0.62\textwidth}{!}{%
  \includegraphics{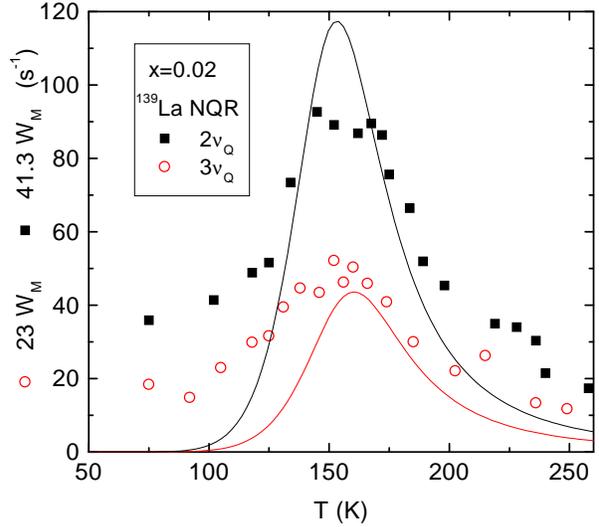}
  }
  \vspace{-1.5cm}       
  \caption{NQR relaxation rate of magnetic origin obtained from the data in Fig.4
  after the subtraction of the quadrupolar background contribution, in LSCO at  
  x = 0.02, for the 3$\nu_Q$  and the 2$\nu_Q$ line. The solid lines are the theoretical
  behaviors  according to Eq. 5 in the text   ( $|h|^2$ as adjustable parameter)
  by using for $\omega_s$ the form deduced from anelastic relaxation ( Eq. 2 in the text ).}
  \label{fig:5}       
  \vspace{-0.6cm} 
 \end{figure}
On the other hand, if the maxima in $W_M$ at 150 K are taken as an 
indication of stripe motion at frequency around $\omega_Q$, then an NQR line
broadening must be expected below the temperature at which the frequency
becomes of the order of the line width itself, about 200 kHz, namely
around 110 K according to Fig. 3. An experimental support to the
hypothesis that the intrinsic linewidth $\delta$  is temperature dependent 
comes from the comparison of the spectra at 2$\nu_Q$ and at 3$\nu_Q$ (Fig.6).
The ratio $\delta_{2\nu_Q}$(T = 77 K) / $\delta_{2\nu_Q}$(T = 177 K) 
= (210 kHz ) / (183 kHz) = 1.15
is the same of the one for the 3$\nu_Q$ line:
$\delta_{3\nu_Q}$(T = 77 K) / $\delta_{3\nu_Q}$(T = 177 K) 
= (320 kHz ) / (280 kHz) = 1.15
One also has $\delta_{3\nu_Q}$ = (3/2) $\delta_{2\nu_Q}$. These data are not compatible with 
an effect due to a magnetic field $\langle h\rangle$, that would cause an extra
broadening of the same amount for both the 2$\nu_Q$ and the 3$\nu_Q$ lines. 
Thus, if a moderate (of the order of 15-20\%) temperature dependence
for the single component linewidth $\delta$  is allowed, then the NQR 
spectra indicate T$_N \simeq$  50 K. This value is in substantial agreement 
with the phase diagram commonly accepted in literature \cite{john}. One could
speculate that at this temperature a small bump is observed in the
relaxation rates (Fig.4), consistent with the slowing-down of the 
spin dynamics, just above the temperature range where the drastic 
increase of $W_M$ due to the spin freezing occurs. On cooling, the
relaxation rate exhibits a maximum at a temperature around T = 9 K.
Also the recovery plots, showing evidence of departure from an
exponential recovery towards the t$^{1/2}$ law, support the conclusion 
that the spin-glass quasi-freezing temperature T$_f$ has been reached.
 \begin{figure}
  \resizebox{0.62\textwidth}{!}{%
  \includegraphics{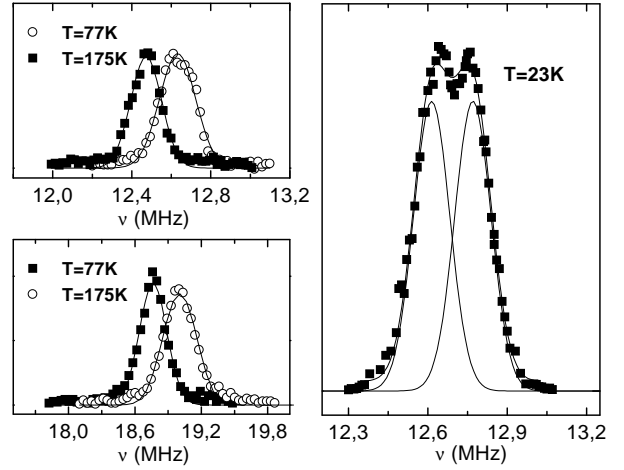}
  }
  \vspace{-1.5cm}       
  \caption{Typical La NQR spectra ( obtained by Fourier transforming the 
  signal as a function of the irradiation frequency) in LSCO at x = 0.02  
  for the 2$\nu_Q$ and the  3$\nu_Q$ lines.}
 \end{figure}
We compare now the experimental findings in the sample at the boundary
between the AF and the spin-glass phase with the one at x = 3 percent,
well within the latter phase. The anelastic relaxation shows a temperature 
behavior similar to the one for x = 2 percent (Fig.7). Both the peaks
attributed to the stripes motion and to collective tilting of the 
octahedra (not shown) are attenuated by a factor of about two. The 
\la NQR relaxation rates indicate the typical freezing of the spin
fluctuations in a cluster spin-glass (Fig.8). The maximum in $W$ 
occurs at T = 8 K, in good agreement with the phase diagram by Cho et al \cite{cho,cho2}. 


The relaxation rate measured at 2$\nu_Q$ reaches a value about twice 
the one reported in previous measurements \cite{cho,borsa2,riga} at 3$\nu_Q$ , 
consistent with a magnetic relaxation mechanism . It is noted 
that for x = 0.03 magnetization measurements provide direct 
evidence of the occurrence of a canonical spin-glass state \cite{waki}. 


The temperature dependence of $W$ can be discussed in terms of
the behavior expected for the effective correlation time.
For magnetic moments coupled to a Fermi gas of carriers \cite{macf}, from Eq. \ref{tau} 
in the fast fluctuations regime one would have $W\propto 1/\omega_s = \tau_f = h/\pi 
(\rho J)^2 kT$, 
where $J$ is the exchange coupling to the band and $\rho$ the density of states at
the Fermi level. From Fig. 8 one sees that $W$ diverges, on decreasing temperature, 
more rapidly than T$^{-1}$, in a way close to the law $\tau \propto exp[E/T]$ at least for the 
temperature range where the fast motions condition, namely $\tau \omega_Q\ll 1$, holds 
(see inset in Fig. 8). Below T$\simeq$ 10K one notes the behaviour of the
 relaxation rate expected for a glassy system \cite{chou,cho}.


\begin{figure}
\resizebox{0.65\textwidth}{!}{%
  \includegraphics{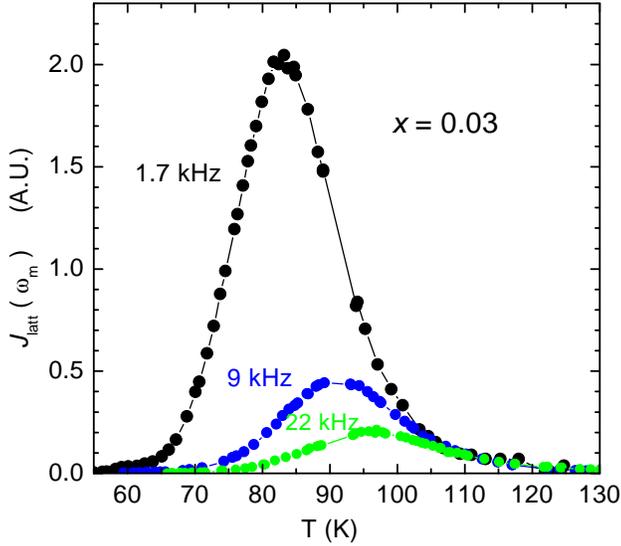}
}
\vspace{-1.5cm}       
\caption{Spectral densities of the motions 
responsible of the elastic energy loss in LASCO at x=0.03, 
for three measuring frequencies.}
\label{fig:7}       
\end{figure}
\begin{figure}
\resizebox{0.58\textwidth}{!}{%
  \includegraphics{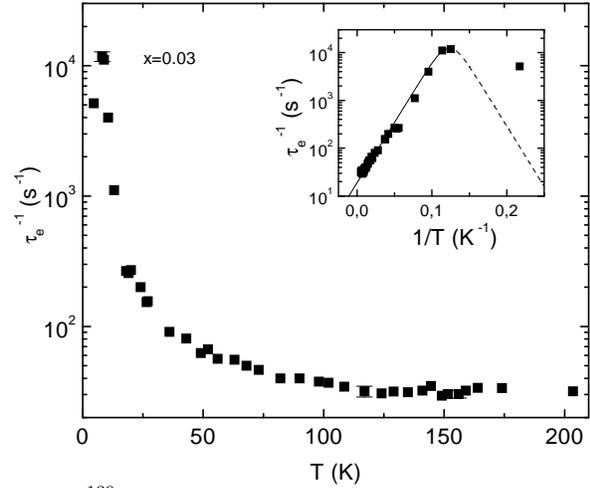}
}
\vspace{-1.5cm}       
\caption{\la NQR relaxation rate for the 2$\nu_Q$ line in LSCO at x =0.03. 
In the inset the data 
 for $\tau_e^{-1}$ are fitted according to the law $\tau_e^{-1}\propto \tau \propto 
exp[E/T]$ with E$\simeq$58 $\pm$ 2 K. }
\label{fig:8}       
\end{figure}


\vspace{0.5cm}
\section{Conclusions}
Anelastic and NQR relaxation and NQR spectra have been combined in the 
attempt to derive insights on spin and lattice excitations, possibly 
stripes motions, driving the ordering processes in LSCO. The experimental 
findings have been discussed within two interpretative frameworks. On 
one side it could be possible that the phase diagram around the Sr 
content x = 0.02 separating the AF and the spin-glass phases is more 
complex that previously assumed, with a quasi-long range ordering 
temperature as high as 150 K, corroborating recent neutron scattering
 measurements \cite{birge} and qualitatively agreeing with the extrapolation 
at x = 0.02 of a magnetization study \cite{waki} carried out in samples at
 x = 0.03, 0.04 and 0.05. The order parameter of such a transition 
would be characterized by an unconventional temperature dependence.


On the other hand, the NQR spectra can be interpreted as indicating
 a conventional transition to the AF phase around T = 50 K, in
 substantial agreement with the phase diagram commonly accepted. 
In this case the anelastic and La NQR relaxation rates around 
T = 80 K and T = 160 K respectively, are the first direct
 experimental evidence of low frequency motions of stripes 
simultaneously involving spin and lattice excitations. 
The thermal depinning barriers and the characteristic 
\`diffusive\' frequencies are then derived, and they do not
 differ much in the sample at Sr content 0.03.


In the low temperature range a spin freezing process is
 detected, with a dramatic increase of the \la NQR 
relaxation rate on cooling. 


These experimental findings could be, at least in part,
 explained with a kind of distribution of transition 
temperatures T$_N$, T$_f$ and T$_g$ resulting from a spread in 
the Sr content around the critical amount x = 0.02.
 However the tetragonal-orthorhombic transition
 appears sharp and hence the Sr content seems to be 
well defined and close to that value.


\section{Acknowledgments}
Alessandro Lascialfari is gratefully thanked for his SQUID 
measurements and for helpful discussions. Stimulating discussions
 with F. Borsa, P. Carretta, R. Gooding and M.H. Julien are also acknowledged.


 The research has been carried out in the framework of the PRA 
project SPIS (1998-2000), financed by INFM (Italy).
%

\begin{thebibliography}{}
\bibitem{ferr} The samples 
have been prepared and characterized by X-ray diffraction by 
M. Ferretti and coworkers (Dep. of Chemistry, University of Genova).
\end{thebibliography}
%



\end{document}